# Electron Localization on Molecular Surfaces by Metal Adsorption

Abraham F. Jalbout[a*] and Thomas H. Seligman[b,c]

[a]*Instituto de Química, Universidad Nacional Autónoma de México, Circuito Exterior, Ciudad Universitaria 04510, México D.F., México*

[b]*Instituto de Ciencias Físicas, Universidad Nacional Autónoma de México, Cuernavaca, Morelos, México*

[c]*Centro Internaciona de Ciencias A.C. Cuernavaca, Morelos, México*

*Abstract:*

The ability of metal adsorption to transfer charge to the surface of single molecular carbon sheets is explored in this paper. Though other metals are considered we basically will deal with Lithium We concentrate on fairly small sheets and examined the minimum threshold size of a molecular surface needed to separate metals. From our quantum chemical calculations we deduce that a molecular surface of six benzene rings is needed for Lithium dimers to be separated. We further observe symmetry breaking, when two lithium atoms are adsorbed right opposite to each other on the two sides of the sheet.

---

*To whom correspondence should be addressed.
E-mails: ajalbout@u.arizona.edu (A.F. Jalbout)

**Introduction**

In past investigations, it has been shown a novel type of electron traps to be of interest. These studies focus on using molecules with internal dipole moments to localize increased electron density and thus charge on molecular surfaces [1-7]. Systems of this type can be useful not only for fundamental research but also in the design of advanced materials.

It has been shown that Li is capable of transferring the equivalent of a full electron to regions of a fullerene molecular surface [8]. Similar results hold for nano-tubes and improve Whose its solubility properties in molecular surfaces [9]. The consequent modification of the electron surface configuration can lead to increased reactivity in specific regions of the molecule.

The purpose of the present work to consider the ability of metals to transfer charge to the surfaces of molecules. We will explore this question by calculating metal adsorption properties in small molecular surfaces composed of benzene rings. Such polyaromatic hydrocarbons (PAH´s) have been shown to exist experimentally [10a], and indeed are readily formed.

Literature research has reveals that studies exist on charge transfer to surfaces [10b] and symmetry breaking [11-12] few deal with electron localization [13]. The primary principle at the origin of the present report is that metal ions have been shown to be incorporated into benzene rings with some affinity [14]. This coupled to the intrinsic ability of Li to transfer electrons to the surfaces of aromatic surfaces has lead us to study the systems mentioned. The effects that we are attempting to compute are much weaker on cyclohexane surfaces due to the fact the partial polarity of the hydrogen atoms which will prevent adequate charge transfer of electrons to the surface by metal adsorption. This has been shown to be a local effect in benzene rings [14] and should be observed in larger molecular surfaces. The calculations therefore serve as benchmarks for how electron localization in particular regions along the molecular surfaces can adequately take place.

**Computational Methods**

The quantum chemical calculations were performed with the GAUSSIAN03 program codes [15]. Due to the relatively large size of the systems, geometry optimizations were performed with the Hartree-Fock (HF) method coupled to the 3-21G* basis set. Higher level calculations require vast computational effort, but since it is our goal to provide general trends and observations the proposed level should suffice. Additionally, we have complemented the calculations with others for sodium (Na) and potassium (K) metals and the results were consistent with those for Lithium. Thus the adsorption of litium (Li) will be used as test case.

**Results and Discussion**

Figure 1 displays the molecular surfaces used and also the structures obtained by single Li metal adsorption to the surface with intermolecular distance displayed (in angstroms, Å). Figure 2 depicts the adsorption of two metal ions to the molecular surface on the same side and opposite sides of the surfaces and the last figure shows the successful separation of $Li_2$ on a six ringed system (**VI**). Table 1 shows the relative adsorption energies (in kcal/mol) for the adsorption of a single Li atom ($\Delta E_1$), the $Li_2$ molecule ($\Delta E_2$), double Li atom adsorption on the same side of the molecular surface ($\Delta E_3$) and on opposing sides ($\Delta E_4$).

The molecular surfaces have been constructed by adding a benzene ring in a configuration that yielded the most favorable interaction with the metals. Several configurations were calculated and we show the most stable ones. For the interaction of a single metal atom with the molecular sheets we can see that the energies of adsorption increases linearly. The first sheet (structure **I**) has an intermolecular separation of around 2.4 angstroms with an adsorption energy of about 24 kcal/mol. A calculation of the partial charge of Li at the HF//6.311G** level of theory using the natural orbital population analysis (NOPA) method yielded a value of 0.83e. This basis set has been shown to be ideal in related studies [10] on Li charge transfer to molecular surfaces.

For the next system (structure **II**) we obtain an intermolecular separation of around 2.4 angstroms with an adsorption energy of 45.4 kcal/mol. System **III**, has a smaller intermolecular separation of around 2.1 angstroms with an energy of interaction of about 46 kcal/mol. We see a similar increase in interaction energies for **IV**, **V** with values of around 72, and 89 kcal/mol, respectively. The **VI** system has a slightly elevated value of 90 kcal/mol for this interaction.

We now move on to double metal adsorption (Figure 2) and see slightly different behavior. While the charge transfer capabilities suggest that also for double adsorbtion the metals transfer the majority of the electron density to the molecular surface, variations in the molecular properties can be observed. For species **I** we can see that the adsorption energy of both Li atoms to the surface is 39 kcal/mol with an intermolecular separation of about 2.3 angstroms and an intramolecular distance of around 2.8 angstroms. The Li-Li bond length is conserved at around 2.8 angstroms in most cases considered and the distance to the surface is around 2.3 angstroms in most complexes.

The next system, structure **II** has an energy of interaction of about 36.5 kcal/mol and **III** has a ΔE of 38.6 kcal/mol. The structure **IV**, **V**, and **VI** have interaction energies of about 36.6, 33.1 and 36.4 kcal/mol, respectively. In all cases except for **III** the intramolecular separation is consistently yielding a value of 3.3 angstroms. For **VI** there is another structure shown in Figure 3 that has an interaction energy of 10.4 kcal/mol and is the dissociation energy of $Li_2$. It represents the minimum threshold value of a molecular surfaces diameter needed to separate two metals from one another. The typical dissociation energy for the isolated Li dimer is around 131 kcal/mol so this represents a significant improvement. This is an improvement since it permits the dimer systems to be separated much more easily on molecular surfaces than in the isolated gaseous state.

The final case considered is the adsorption of two Li atoms to opposing sides of the surface. Again, the charge transfer of the Li atoms to the surface is rather consistent. The first structure has an interaction energy of 19.3 kcal/mol with an intermolecular separation of around 2 angstroms. The next structure shows a significant increase to around 55 kcal/mol with a very similar distance to the surface. This increase can be due to the fact the that the system delocalizes the excess electron density quite well leading to improvements in the electron acceptor properties. For **III**, **IV** and **V** we obtain energies of interaction of 21.4, 22.7, and 11.5 kcal/mol, respectively. In **VI** the interaction energy is around 11.1 kcal/mol. In the latter structures (species **V**, **VI**) the spontaneous symmetry breaking leads to increases in energy which destabilize the interaction. Additionally as the sheet gets larger the metals tend to localize in a specific region of the surface. Therefore, it does not matter how large the systems are they should behave similarly.

**Conclusions**

The present work has explored the adsorption of Li to a surface of a series of small polyaromatic hydrocarbon sheets. The surfaces used were selected since they are highly aromatic due to the presence of the benzene rings. As in fullerenes and nanotubes the aromaticity of the benzene ring components creates a electrostatic field that increases the delocalization of the electrons in the molecular complexes. This aromaticity permits the adequate adaption of the electrons of Li to be transferred to the molecular surface. The Li atom contributes to the electronic density on localized regions of the molecular surface. We have performed test density functional theory (DFT)-B3LYP/6-31+G* calculations and similar qualitative results were observed. We are thus confident that the calculations are sufficiently accurate and qualitatively correct.

These computations effectively show how electrons can be transfer to the surface on localized regions in small systems From the previous computations it has been shown that in single metal systems, the benzene ring adequately accepts excess electrons from the metal into its electron clouds of delocalized electrons. Therefore, the metal to surface charge transfer in extended systems is a logical extension on what has been proposed in single ringed systems [14].

In addition to proposing a methodology for localizing electrons on specialized regions of a molecular surface, we report the minimum threshold size of a molecular surface needed to separate alkali metal dimers. Another interesting property is the symmery breaking when the metals are absorbed on opposite sides of the molecular surface. This is more prolific in smaller systems whereby the Li atoms tend to occupy similar regions on the molecular surface. We believe that this work can be instructive for the analysis of charge transfer in larger molecular frameworks, and a corresponding study is under way.

**Acknowledgments**

Special thanks are extended to DGSCA as well as UNAM for valuable resources. One of us (THS) acknowledges support from PAPIIT-UNAM contract IN112307.

## Table and Figure Captions

**Table 1.** Relative energies (in kcal/mol) for the adsorption of a single Li atom ($\Delta E_1$), the Li$_2$ molecule ($\Delta E_2$), double Li atom adsorption on the same side of the molecular surface ($\Delta E_3$) and on opposing sides ($\Delta E_4$).

**Figure 1.** Basic molecular surfaces and single Li atom adsorption on the molecular surfaces with intermolecular distances shown in angstroms (Å).

**Figure 2.** Double Li atom adsorption on the molecular surfaces with intermolecular distances shown in angstroms (Å).

**Figure. 3.** Double Li atom adsorption on **VI** in the separated case with intermolecular distances shown in angstroms (Å).

| Species | $\Delta E_1$ | $\Delta E_2$ | $\Delta E_3$ |
|---------|--------------|--------------|--------------|
| I   | 23.98 | 39.41 | 19.30 |
| II  | 45.40 | 36.52 | 55.35 |
| III | 46.02 | 38.60 | 21.36 |
| IV  | 71.81 | 36.61 | 22.73 |
| V   | 89.42 | 33.06 | 11.47 |
| VI  | 90.07 | 36.41 | 11.13 |
|     |       | 10.43[a] |    |

**Table 1.**

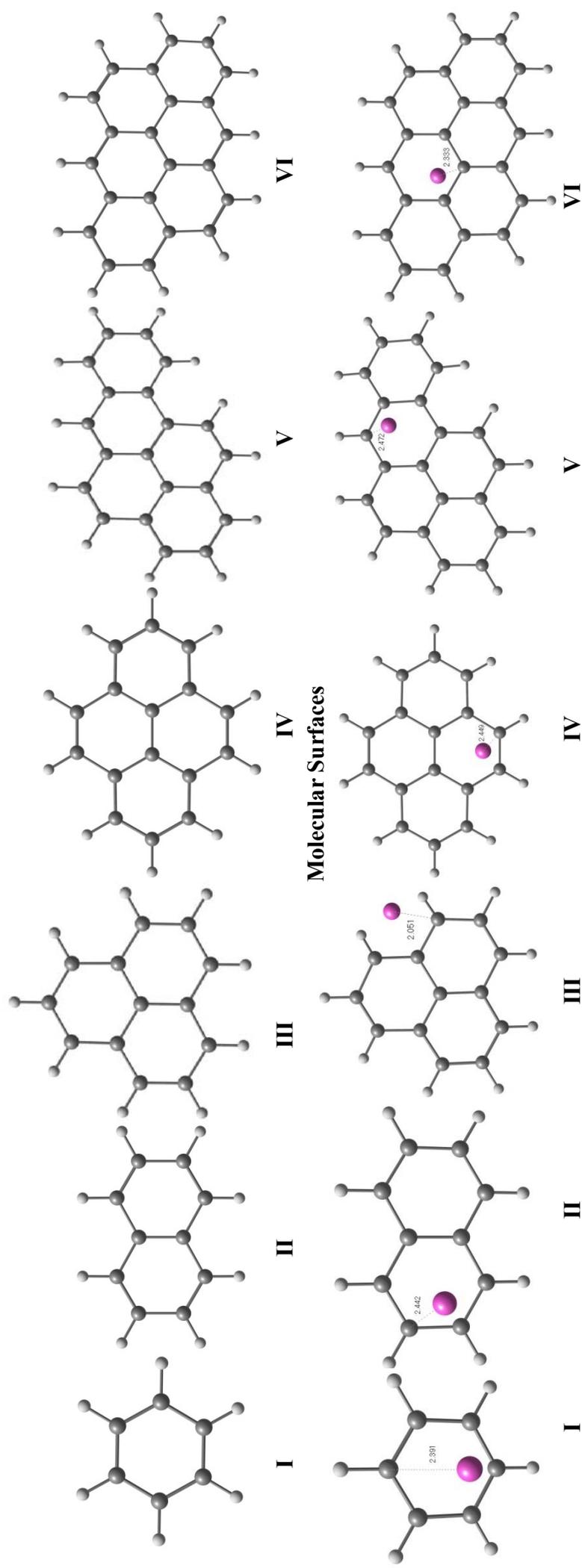

**Figure 1.**

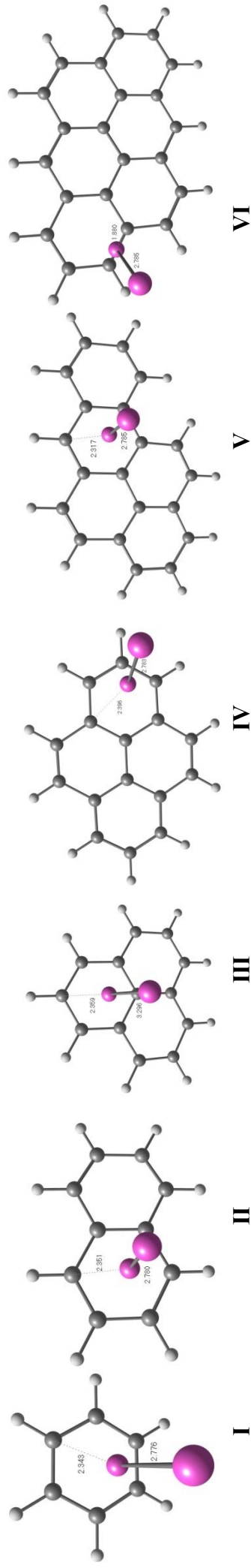
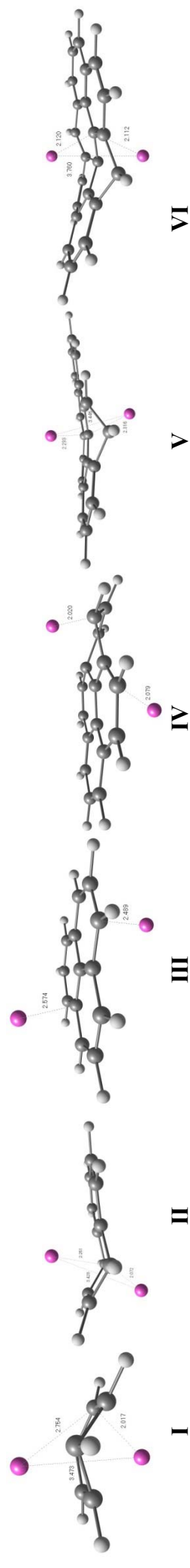

**Figure 2.**

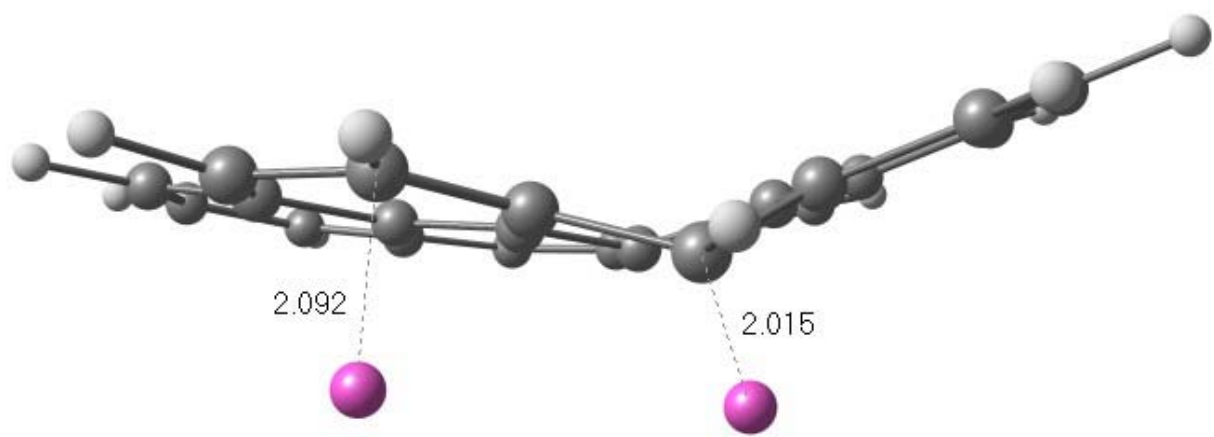

**Figure 3.**